\documentclass[10pt, conference]{IEEEtran}
\IEEEoverridecommandlockouts
\usepackage{cite}
\usepackage{amsmath,amssymb,amsfonts}
\usepackage{algorithmic}
\usepackage{graphicx}
\usepackage{svg}
\usepackage{textcomp}
\usepackage{subcaption}
\usepackage{comment}
\usepackage{adjustbox}
\usepackage{xcolor}
\usepackage[ruled,lined,boxed,linesnumbered,commentsnumbered,longend]{algorithm2e}
\usepackage{optidef}
\usepackage{multicol}
\usepackage{nopageno}
\makeatletter 
\newcommand{\linebreakand}{%
  \end{@IEEEauthorhalign}
  \hfill\mbox{}\par
  \mbox{}\hfill\begin{@IEEEauthorhalign}
}
\makeatother 

\def\BibTeX{{\rm B\kern-.05em{\sc i\kern-.025em b}\kern-.08em
    T\kern-.1667em\lower.7ex\hbox{E}\kern-.125emX}}
\begin{document}

    \title{Towards Energy Efficiency in RAN Network Slicing
}
\author{\IEEEauthorblockN{Hnin~Pann Phyu}
\IEEEauthorblockA{Département de Génie Logiciel et TI \\
École de Technologie Supérieure (ÉTS)\\
Montreal, Canada \\
hnin.pann-phyu.1@ens.etsmtl.ca}
\and
\IEEEauthorblockN{Diala Naboulsi}
\IEEEauthorblockA{Département de Génie Logiciel et TI \\
École de Technologie Supérieure (ÉTS)\\
Montreal, Canada \\
diala.naboulsi@etsmtl.ca}
\linebreakand
\IEEEauthorblockN{Razvan Stanica}
\IEEEauthorblockA{Univ Lyon, INSA Lyon, Inria, CITI \\
Villeurbanne, France \\
razvan.stanica@insa-lyon.fr}
\and
\IEEEauthorblockN{Gwenael Poitau}
\IEEEauthorblockA{ Dell Technologies \\
Montreal, Canada\\
gwenael.poitau@dell.com}
\thanks{This work was supported by the National Natural Sciences and Engineering Research Council of Canada (NSERC) through research grant RGPIN-2020-06050 and by the CHIST-ERA ECOMOME project, through the Fonds de Recherche du Québec – Nature et Technologies (FRQNT).}
}

\maketitle
\vspace*{-5mm}
\thispagestyle{plain}
\pagestyle{plain}

\begin{abstract}
Network slicing is one of the major catalysts to turn future telecommunication networks into versatile service platforms. Along with its benefits, network slicing is introducing new challenges in the development of sustainable network operations. In fact, guaranteeing slices requirements comes at the cost of additional energy consumption, in comparison to non-sliced networks. Yet, one of the main goals of operators is to offer the diverse 5G and beyond services, while ensuring energy efficiency. To this end, we study the problem of slice activation/deactivation, with the objective of minimizing energy consumption and maximizing the users quality of service (QoS). To solve the problem, we rely on two Multi-Armed Bandit (MAB) agents to derive decisions at individual base stations. Our evaluations are conducted using a real-world traffic dataset collected over an operational network in a medium size French city. Numerical results reveal that our proposed solutions provide approximately 11-14\% energy efficiency improvement compared to a configuration where all the slice instances are active, while maintaining the same level of QoS. Moreover, our work explicitly shows the impact of prioritizing the energy over QoS, and vice versa.




\end{abstract}


\section{Introduction}\label{intro}

The telecommunication industry accounts for approximately 2\% of total global carbon emissions~\cite{Cubukcuoglu2022}. By 2030, 8\% of the projected global electricity demand will come from the information and communications technology sector as a whole, even in the best case scenario~\cite{Jones2018}. Energy consumption will continue increasing in beyond 5G and 6G networks, where computationally intensive services will be largely deployed. Although 5G equipment is more energy efficient than 4G~\cite{Erfanian}, with the data traffic volume increasing tremendously along with 5G services, overall energy consumption will increase too. In fact, the energy consumption of a 5G base station is three times higher than that of a 4G base station, when both are considered at a full load~\cite{Fan2022}. 

5G is envisioned to serve a wide variety of services, with heterogeneous traffic, through network slicing~\cite{Foukas2017a}. This is done by forming, on one physical network, multiple virtual networks on a per-service basis, i.e., slices. That said, slices requirements need to be met, including performance isolation. Guaranteeing these requirements and the additional virtualization layer come with some overhead, which produces higher energy consumption with respect to non-sliced network deployments~\cite{Masoudi2022}. One of the key objectives in the field is to offer this service differentiation, while reducing the associated CO$_2$ emissions. Indeed, energy efficiency in networks is no longer an option but a necessity. When delving into this topic, we observe that, today, the highest amount of energy is consumed in the radio access network (RAN), approximately 70\% of the overall network energy utilization~\cite{Piovesan2022}. 

To deal with this, several research works consider base station sleep schemes to further optimize the energy consumption in 5G networks~\cite{Han2013, Feng2017}. While such techniques show effective results, they are more challenging to be applied directly in the case of multi-services network slicing environments. That is mainly because slice instances can exhibit quite different temporal traffic patterns. Completely shutting down or putting the entire base station into sleep mode could notoriously impact the quality of service (QoS) of users in specific slice instances. This motivates us to introduce a new approach, in which slice instances are dynamically activated and deactivated, according to their traffic patterns, thereby enhancing the overall base station energy efficiency. However, deactivating some slice instances to minimize the energy consumption can potentially degrade the QoS of users. On the other hand, activating all slices all the time, so as to maximize QoS, significantly increases energy consumption. Accordingly, the energy minimization objective shall be coupled with a QoS maximization objective~\cite{Chatzipapas2011}.


To manage the trade-off between the two objectives, operators may consider using an EcoSlice, which is a slice instance with bare minimum resources and network functions. By that, it incurs much lower energy consumption than typical slice instances. The EcoSlice is up and running $24/7$ to provide a bare-minimum service. In some conditions, e.g., low traffic demand, operators may switch the users of other slices to this specific EcoSlice, without a significant QoS impact.


In this regard, we study the problem of slice activation/deactivation, with the objective of minimizing the energy consumption while satisfying the user QoS. The contributions of our work are twofold. First of all, we propose two different approaches for solving the problem, namely a Deep Contextual MAB (DCMAB) algorithm and a Thompson Sampling Contextual (Thompson-C) algorithm. These approaches allow to derive solutions dynamically over time, while considering traffic patterns of individual slice instances deployed at a base station. Moreover, our proposed agents enable operators to navigate users to and from an EcoSlice, if their requested slice instance is activated/deactivated. Second, we evaluate the performance of the proposed approaches and their computational cost using a real-world traffic dataset. 

The rest of the paper is organized as follows. Section~\ref{sect2} discusses the related work of energy efficiency in network slicing. Then, the network model and problem statement are laid out in Section~\ref{sect3}. In Section~\ref{sect4}, we present the detailed design of our proposed solutions. We articulate the results in Section~\ref{sect5} and conclude the paper in Section~\ref{sect6}. 

\section{Related Work} \label{sect2} 

With the aim of enabling energy efficiency, several research works consider optimizing the allocation of network slice resources (i.e., radio, CPU, transmission bandwidth and power) in the different domains (i.e., RAN, Edge Computing (EC), Core Network (CN) and end-to-end network). At the RAN slicing level,~\cite{Azimi2022} combines deep learning (DL) and reinforcement learning (RL) on a distributed framework to efficiently allocate radio and transmission power resources over base stations. They use stacked and bidirectional Long-Short Term Memory (SBiLSTM) to predict the per slice resources demand on a large time scale and rely on asynchronous advantage actor-critic (A3C) to allocate resources to users on a small time scale. Their proposed framework achieves higher energy efficiency than baselines using static power allocation. 

In~\cite{Rezazadeh2021}, the authors optimize the energy consumption and computation cost in a network slicing based Cloud-RAN (C-RAN) setting, using a twin-delayed double-Q soft Actor-Critic (TDSAC) approach. Their agent performs the up/down scaling of computing and beamforming power resources. Their work outperforms other baseline RL models in terms of overall network energy and computing cost. Similarly,~\cite{Masoudi2022} designs a slice energy consumption model based on the C-RAN architecture. An optimisation problem is solved per-slice, with the objective of minimizing the overall network energy cost, jointly considering communication and computation resources. This approach improves energy efficiency over a baseline focusing only on radio resources. 


Focusing on CN slicing, the authors in~\cite{Akin2022} formulate a security-aware network slicing optimization problem to enhance the energy efficiency of CN nodes. They limit themselves to static resource allocation. Their proposed solution provides more power savings than a greedy approach. 


Considering an obvious trade-off, it is sensible to couple QoS maximization and energy consumption minimization. Therefore, focusing on end-to-end network slicing, the authors in~\cite{Chergui2021} aim to maximize the energy efficiency while respecting service level agreement (SLA) constraints. To this end, they rely on statistical federated learning (stFL). Their federated local agents coordinate and predict per slice network metrics, without transferring datasets to a central unit, and largely outperform other federated learning and centralised solutions. 



While prior works attempt to achieve the energy efficiency as well as ensuring QoS, we believe there is still room to further optimize the energy efficiency by switching off some of the underutilized slice instances, under some conditions. As of our knowledge, there is no contemporary work studying this problem. In this light, we introduce the RAN slice activation/deactivation problem with the aim of minimizing energy consumption and maximizing QoS. To this end, we rely on the fully decentralized state-aware MAB approaches to enable decisions for slice instances at each base station, while considering the impact on energy and QoS factors.

\section{System Model and Problem Statement} \label{sect3}

We study the problem of slice instance activation and deactivation at individual base stations with the aim of minimizing energy consumption and maximizing QoS. We thus explicitly lay out the system model, energy consumption and user QoS model, needed as part of our defined problem. We then formalise the main objective of our problem. We formulate the latter as a Markov decision process (MDP), and re-design it after that as a contextual MAB problem. 

\subsection{System Model} \label{IIIA}

\begin{figure}[ht!] 
\vspace{-0.25cm}
    \centering
    \includegraphics[scale=0.3]{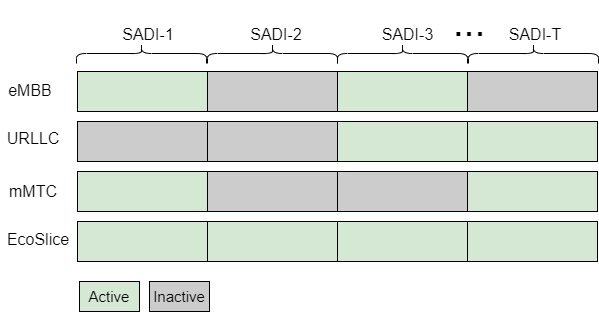}
    \caption{Slice Activation-Deactivation Interval (SADI) illustration.}
    \label{sadi}
\end{figure}

We consider a time-slotted system. Accordingly, we define $\tau$ as the slice activation/deactivation interval (SADI), where slices are active or inactive continuously over the period of that interval. Accordingly, activation/deactivation decisions are made at the end of every $\tau$, for the upcoming $\tau +1$. The period of the SADI is defined based on the operator policies. Besides, we define $T$ as a time interval of interest, such that $\tau \in T$. Figure~\ref{sadi} illustrates the time frame consideration in our proposed model. In this example figure, we consider four different types of slices: three of them, denoted as Enhanced Mobile Broadband (eMBB), Ultra-Reliable Low Latency Communication (URLLC) and Massive Machine-Type Communication (mMTC), provide services with different QoS levels and they can be activated/deactivated as needed, as well as the EcoSlice which is always up and running.


We define $\mathcal{I}_{b}$ as a set of slice instances attached to base station $b \in \mathcal{B}$. We define $\mathcal{U}_{b}^{\tau}$ as the set of users served by base station $b$ at $\tau$. Accordingly, $U_{i,b}^{\tau}$ is the set of users that can be served by slice instance $i$ of base station $b$ at $\tau$. Thus, $\mathcal{U}_{b}^{\tau} = \bigcup_{i \in \mathcal{I}_{b}}U_{i,b}^{\tau}$. Each slice instance $i \in \mathcal{I}_{b}$ is characterized by a specific QoS class identifier (QCI)~\cite{Sun2019} and its energy consumption $E_{i,b}^{\tau}$. 

Without loss of generality, we consider user delay as an indicator of the QoS, but any other metric could be easily integrated instead. In the optimal slice instance activation scheme, underutilized slice instances are switched off to save energy when certain conditions are met. Consequently, the user-perceived delay is prone to deteriorate. In this light, we deliberately study the impact of optimal slice instance activation on both energy and delay. Hence, we define $\delta_{i}$ as a predefined achievable delay for a slice instance $i$. We also consider an EcoSlice instance $i_{e} \in \mathcal{I}_{b}$, to which users are switched when their requested slice instances are inactive. The EcoSlice instance $i_{e} $ also has a predefined achievable delay $\delta_{i_{e}}$ and its energy consumption over ${\tau}$ is denoted as $ E_{i_{e},b}^{\tau}$.

At every $\tau$, each user $u \in \mathcal{U}_{b}^{\tau}$ makes a specific request, characterized by a delay requirement $d_{u}^{\tau}$ and a traffic flow demand $l_{u}^{\tau}$. Each base station $b$ has a set of possible configurations $\mathcal{K}_b$. Each configuration $k \in \mathcal{K}_b$ implies activation/deactivation decisions for some slices. In detail, $k$ contains $\left\{ c_{i}^{\tau} |  i \in \mathcal{I}_{b} \right\}$. Here, $c_{i}^{\tau} = 1$ if slice instance $i$ is active at $b$ during $\tau$, and 0 otherwise. Needless to say, $c_{i_{e}}^{\tau} = 1$ for any $\tau$.

\subsection{Energy Consumption Model}

We define the function $f(.)$ to refer to the overall energy consumption of a base station. This energy model can be further fine-tuned based on specific operator resource management policies and activated RAN energy saving features. It is composed of the energy consumption resulting from its individual deployed slice instances $E_{i,b}^{\tau}$ and the static energy consumption of the base station $P_{b}^{static}$ (i.e cooling and circuit power):
\begin{equation}
 f_{b}( c_{i}^{\tau}, \mathcal{I}_{b}  )= \sum_{i \in \mathcal{I}_{b}} c_{i}^{\tau} \cdot E_{i,b}^{\tau} + P_{b}^{static} \label{eq1}
\end{equation}
with
\begin{equation}
E_{i,b}^{\tau} = \rho_{i,b}^{\tau} \psi_{i} P_{b}^{dynamic}+  \psi_{i} P_{b}^{fixed} \label{eq2}
\end{equation}

As indicated in the above equation, the energy consumption of slice instances for a base station $b$ encompasses a load-dependent power consumption component, $P_{b}^{dynamic}$, and a load-independent power consumption component, $P_{b}^{fixed}$. Specifically, as the name implies $P_{b}^{dynamic}$ depends on the traffic load of the base station. It is worth stressing that a slice instance consumes some power even at zero load traffic, in order to run its corresponding network functions. Thereupon, $P_{b}^{fixed}$ is independent of the traffic load, but related to the energy consumption of associated network functions of specific slice instances. 

We note here that not all slice instances are designed equally~\cite{Marquez2017}. Their required network functions and signalling traffic are different~\cite{Guan2018}. For instance, an URLLC service potentially consumes higher energy, because it requires specific network functions to offer high reliability and very low latency~\cite{Bairagi2021}.  In short, the more stringent the latency requirements of the service, the higher its consumed energy~\cite{Bennis2018}. Regardless of their requirements for energy, both mMTC and eMBB have flexibility in terms of latency. It is fair to say that, even under the same amount of traffic load, the energy consumption of each service is different. Meanwhile, the EcoSlice is deployed with relatively low energy consumption. It is sensible to conclude that different slice types consume different amounts of energy not only because of their traffic portion but also because of their service attributes~\cite{Marquez2019}.

In this vein, $\psi_{i}$ denotes the power consumption impact factor of slice instance $i$ on both $P_{b}^{dynamic}$ and $P_{b}^{fixed}$. We let the operators define the value of $\psi_{i}$ based on their slice instances configurations and analytics. Having said that, based on the prior explanation, it is pragmatic to assume that URLLC has larger $\psi$ value than eMBB and mMTC. Needless to say, $\psi$ value of the EcoSlice is the lowest. Besides, $\rho_{i,b}^{\tau}$ is the traffic load portion of associated slice instance $i$ of base station $b$ during ${\tau}$. For this, we can simply obtain $\rho_{i,b}^{\tau}$ by dividing the traffic load over slice  $l_{i,b}^{\tau}$ by the total base station traffic load, as below:
\begin{equation}
    \rho_{i,b}^{\tau} = \frac{l_{i,b}^{\tau}}{\sum_{i\in I_{b}} l_{i,b}^{\tau}} \label{eq3}
\end{equation}


\subsection{User QoS Model}

As explained, our objective is coupled with ensuring the user QoS. In this regard, we define the user satisfaction factor $\eta_{u}^{\tau}$ for each user $u$ at $\tau$ associated to slice instance $i$ as follows:
\begin{equation}
    \eta_{u}^{\tau}=  \begin{cases}
1 & \text{if}  \ \delta_{i} \leq  d_{u}^{\tau} \\
 0& \text{ otherwise }
\end{cases}, u \in U_{i,b}^{\tau}, i  \in \mathcal{I}_{b}, \tau \in T \label{eq4}
\end{equation}
where $d_{u}^{\tau}$ is the delay requirement of user $u$ at time $\tau$. Consequently, the average per slice QoS at base station $b$ is defined as:
\begin{equation}
\eta_{i,b}^{\tau}= \frac{\sum_{u \in U_{i,b}^{\tau}} \eta_{u}^{\tau}} {|U_{i,b}^{\tau}|} \label{eq5}
\end{equation}

We then compute the average QoS of the base station $b$ considering all the associated slice instances during $\tau$:
\begin{equation}
\eta^{\tau}_{b}= \frac{\sum_{i \in \mathcal{I}_{b}}  \eta_{i,b}^{\tau}}{|\mathcal{I}_{b}|} \label{eq6}
\end{equation}

\subsection{Objective Function}

Given the system model and utility models mentioned in the preceding sections, the objective of our slice activation/deactivation problem can be expressed as:

\begin{equation}
max \  \sum_{\tau=1}^{T}  \left[ \frac{1}{ f_{b}( c_{i}^{\tau}, \mathcal{I}_{b})} +  \eta^{\tau}_{b}\right]   \label{eq7a}
\end{equation}

As one can see in Equation~\ref{eq7a}, the objective function is influenced by the time-varying user demand and active slice instances. Thus, we believe that RL-based approaches are best-suited for this problem, as they enable complex decision-making without requiring an explicit modeling of the network environment~\cite{10103689}. In what follows, we formulate the problem as MDP and contextual MAB.

\subsection{Markov Decision Process (MDP)}

An MDP is defined by a tuple $(\mathcal{S},\mathcal{A}, P, \mathcal{R})$. $\mathcal{S}$ denotes the set of states in the system. Representing states as feature vectors helps the RL agent converge to a near-optimal reward value, but one can also simplify a problem with a less complex state representation (which might converge to a similar reward, with less computation). In this light, we explore two definitions for a state $s$ in this problem: \emph{i)} energy consumption and QoS of base station in the previous SADI: $s= \left\{  f_{b}( c_{i}^{\tau-1}, \mathcal{I}_{b}  ) , \eta^{\tau-1}_{b} \right\}$ and \emph{ii)} simply the SADI identifier: $s=\left\{ \tau \right\}$. The reason for the latter definition is that, since the user demand depends on time and it shows a significant periodicity, a simple state representation based only on the SADI might already contain enough information for the RL agent. Each action $a \in \mathcal{A}$ is a configuration $k$, as defined in Section \ref{IIIA}, and $\mathcal{A}$ is the set of possible configurations: $\mathcal{A}=\mathcal{K}_b$.


In the MDP model, $P: \mathcal{S} \times  \mathcal{A} \times  \mathcal{S} \rightarrow \left[0,1 \right]$ captures the stochastic transition probability function to transition to $s'$ from state $s$ based on action $a$, with $\sum_{s' \in \mathcal{S}} P(s'| s, a)=1$  for all $s \in \mathcal{S}$ and $a \in \mathcal{A}$. $P$ is unknown to an RL agent. However, it occurs that if an agent knows the current state and the reward obtained in each iteration, it can still converge to the optimal solutions through RL approaches~\cite{Shoham2009}. 
Accordingly, the formulation of reward function is critical for the RL agent to be able to learn the optimal policy and it usually boils down to the main objectives of the problem. We therefore define reward $r \in \mathcal{R} $ as: 
\begin{equation}
r(s,a,s') =  \frac{1}{ f_{b}( c_{i}^{\tau}, \mathcal{I}_{b})} +  \beta \cdot \eta^{\tau}_{b}  \label{eq8}
\end{equation}

As shown in Equation~\ref{eq8}, our reward function is aligned with our objective function. Besides, we define $\beta$ as a QoS impacting factor on the reward. In short, the larger the $\beta$ value is, the more the QoS is emphasized with respect to the energy consumption of the base station. If $\beta=1$, the objective is to find the trade-off between energy and QoS.

Due to the nature of our defined problem, we observe two key points in our MDP representation: \emph{i)} the transition probability of MDP can be simplified, such as $P(s'|s,a) \equiv  P(s'|a)$, where states are independent of each other, and \emph{ii)} unlike typical MDP~\cite{Rezazadeh2021}, our reward function depends only on current state and action, but not on the successor state. With this, the reward function could be simplified as $r(s, a, s') \equiv  r(s, a)$. Based on these observations, we present an equivalent formulation as a state-aware MAB in the following.

\subsection{Multi-armed Bandit (MAB)}

In this section, we formulate our slice instance activation/deactivation problem as a state-aware MAB. Formally, state-aware MAB is tupled with $(\mathcal{S}, \mathcal{A}, \mathcal{R})$. Same as in our MDP model, we use the two different state definitions for $s \in \mathcal{S}$: energy consumption and QoS observed over a SADI, or simply the SADI identifier. Similarly, the set of actions $a \in \mathcal{A}$ is the set of available configurations.  

We then define the associated reward set $ \mathcal{R} $. Since we consider the state-aware MAB (i.e., involving multiple states), our reward distribution is non-stationary, and changes based on the state $s$ (also called context in the following). With this, the reward set can be defined as $ \mathcal{R}=\left\{r (s ,a)|\  a\in \mathcal{A}, s \in \mathcal{S} \right\}$. In this regard, we rely on the same reward calculation as Equation~\ref{eq8}. Needless to say, the objective here is to maximize the expected reward $ E \left[ \sum r(s,a)\right]$.

To evaluate our reward function, one standard approach is to compete with the best-action benchmark. On the other hand, we compute the regret resulting from not selecting the optimal action at each iteration. That said, one would define the cumulative regret incurred by an agent over a total of $J$ time steps as~\cite{Slivkins2019}:
\vspace{-0.3cm}
\begin{equation}
Regret (J) =   \sum_{j=1}^{J} ( r_{j}^{*} (s,a)  -  r_{j}(s,a_{j}))
\end{equation}
where $r_{j}^{*} (s,a)$ is the best-action benchmark at round $j$ and can be obtained via  $r_{j}^{*} (s,a) = \underset{a}{max} \ r_{j}(s,a)$, and $a_{j}$ is the action selected by the agent in round $j$. It is worth mentioning that the regret function is non-negative, as it compares the optimal reward to the actual reward obtained by the agent. 

\section{Proposed Solution} \label{sect4}
In this section, we lay out our two fully decentralized approaches, based on DCMAB and Thompson-C agents.

\subsection{DCMAB Agent}

As explained, we use two different context/state definitions and thus we have two types of DCMAB agents: \emph{i)} DCMAB-EQ where the state is the overall energy and QoS over the base station: $s= \left\{  f_{b}( c_{i}^{\tau-1}, \mathcal{I}_{b}  ) , \eta^{\tau-1}_{b} \right\}$, and \emph{ii)} DCMAB-SADI when the state is the SADI: $s=\left\{ \tau \right\}$. Due to space limitation, we outline them together in Algorithm~\ref{algo1}. However, we make sure the main differences (occuring at Line 3 and Line 11 of Algorithm~\ref{algo1}) are clearly outlined.

\begin{algorithm} [h]
    \SetKwFunction{isOddNumber}{isOddNumber}
    \SetKwInOut{KwIn}{Input}
   \KwIn{ Probability of selecting a random action $\epsilon$, learning rate $\alpha$, maximum time steps $J$}
   \KwOut{$\widehat{R}(\tilde{w})$}
    Initialize $\tilde{w}$ randomly and $j= 0$\\
    \Repeat {$j > J$}{
    Observe context/state: $s$ - based on state definition \textit{$s= \left\{ f_{b}( c_{i}^{\tau-1}, \mathcal{I}_{b} ), \eta^{\tau-1}_{b} \right\} $ for DCMAB-EQ \textbf{or} $s=\left\{ \tau \right\}$ for DCMAB-SADI}  \\
    Predict Reward Distribution for each action: $\left [  \widehat{R}_{\tilde{w}}(a|s)\right ]_{a\in A}$ \\
    \eIf{\text{generate random probability:} $rand() \ < \ \epsilon$} {
    Select a random action $a$\\
    } {
    Select $a= \underset{a \in A}{argmax} \left [\widehat{R}_{\tilde{w}}(a|s)\right]$\\
    }
    Evaluate reward $r(a| s)$  \\ 
    Update new state $s'$  \\
    Calculate the loss: $\mathcal{L}(\tilde{w})\overset{\Delta}{= }(r(a| s)-\left [  \widehat{R}_{\tilde{w}}(a|s)\right ]_{a})^2$ \\
    Update the weights: $\tilde{w}\leftarrow \tilde{w}- \alpha \bigtriangledown \mathcal{L}(\tilde{w})$ \\
     $j \gets j+1$\\}
    \caption{$\text{DCMAB-EQ and DCMAB-SADI}$}\label{algo1}

\end{algorithm}

The inputs of the DCMAB agent consist of the probability of selecting a random action $\epsilon$, the learning rate $\alpha$ for the deep neural network (DNN) model, and maximum time steps $J$ to train the DCMAB agent. The output is a trained model, which can predict a reward distribution $\widehat{R}(\tilde{w})$ for the available actions of the associated states. With this, the algorithm begins by the initialization of the weights $\tilde{w}$ with arbitrary values and the variable $j$ referring to an iteration and starting with zero (Line 1). At each step, the agent observes a context $s$: for DCMAB-EQ , $s$ is the overall energy consumption and QoS factor and for DCMAB-SADI, $s$ is a SADI identifier (Line 3). Then, it predicts the reward distribution for all the actions (Line 4) for a given context. After that, an action is selected by considering the exploration and exploitation tradeoffs (Line 5 - Line 9). Precisely, the random action is selected with probability $\epsilon$, and otherwise the action giving the maximum reward is selected. Next, the chosen action is applied to the defined network environment, which, after the potential reconfiguration, returns an actual reward (calculated using Equation~\ref{eq8}) (Line 10). Accordingly, the new state $s'$ is obtained based on the action $a$ for DCMAB-EQ, or new state $s'$ is simply the next SADI (which is independent of the previous action $a$) for DCMAB-SADI~(Line 11). Then, the loss between the predicted reward and the actual reward is computed for the purpose of model training (Line 12). We rely on a gradient descent method to update the weights matrix of the DNN with a learning rate $\alpha$ (Line 13). Then, we go for another iteration (Line 14) and the above process is repeated for a maximum number of time steps $J$ (Line 15).

\begin{algorithm} [h]
    \SetKwFunction{isOddNumber}{isOddNumber}
    \SetKwInOut{KwIn}{Input}
    \KwIn{ Context data $\mathbb{C}=(d_{k})_{|T| \times |\mathcal{K}_{b}|}$, number of dimensions of context/state vector $z=|T|$, $\varphi $, $M$, maximum time steps $J$, $\sigma = M \sqrt{9 z ln(\frac{J}{\varphi })}$  }
   \KwOut{$ \mathcal{N} \left( \hat{\mu}  , \sigma^2 D^{-1} \right)$}
    Initialize $D = \mathbb{I}_{z}$ , $\hat{\mu} = (0)_{z }$ , $j=0$\\
    \Repeat {$j > J$}{
     Sample $\tilde{\mu}$ from Gaussian distribution $ \mathcal{N} \left( \hat{\mu}, \sigma^2 D^{-1} \right)$\\
     Select action $a= \underset{k}{argmax} \ d_{k}^{\mathbb{T}} \tilde{\mu}$ \\
     Observe reward $r_{a} $ \\
     $D = D + d_{a} d_{a}^{\mathbb{T}}$ \\
     $\hat{\mu}  = D ^{-1} \left( d_{a} r_{a}  \right)$\\
     $j \gets j+1$\\
     
    }
    \caption{\text{Thompson-C}}\label{algo2}

\end{algorithm}

 \subsection{Thompson-C Agent}
Unlike DCMAB, the Thompson-C agent adopts a statistical approach with the goal of achieving a proper estimation of the posterior distribution of expected reward for each action. The Thompson-C agent (Algorithm~\ref{algo2}) operates as follows. We note that we only consider the SADI identifier as context/state for the Thompson-C agent. Accordingly, the inputs of the algorithm are the context data for all the actions: $\mathbb{C}=(d_{k})_{|T| \times |\mathcal{K}_{b}|}$, where $d_{k} $ is a context vector for configuration $k$, the number of dimensions of the context vector $z=|T|$, the tunable parameters $\varphi$ and $M$ (that can be tuned by the operator as needed) and the maximum number of time steps $J$ to run the Thompson-C. The output is a posterior distribution $\mathcal{N} \left( \hat{\mu}  , \sigma^2 D^{-1} \right)$ of having the optimal parameter $\hat{\mu}$. For better understanding, $\hat{\mu}$ can be regarded as a weighted vector for a z-dimensional context/state. The parameter $\sigma$ can be obtained via $\sigma = M \sqrt{9 z ln(\frac{J}{\varphi })}$.

The algorithm begins by setting the parameter $D$ as the identity matrix with a z-dimensional vector. $\hat{\mu}$ is initialized with zeros as a z-dimensional vector and $j=0$ (Line 1). In each time step $j$ (Line 2), the Thompson-C agent samples a parameter $\tilde{\mu}$, from the posterior distribution $\mathcal{N} \left( \hat{\mu}  , \sigma^2 D^{-1} \right)$ (Line 3). It then selects an action that yields the best sample (Line 4) and observes an associated reward (Line 5). Finally, the parameters $D$ and $\hat{\mu}$ are updated (Line 6 and Line 7). Then, we go for another iteration (Line 8). The above process is repeated for a maximum number of time steps $J$ (Line 9).

\section{Evaluation} 
\label{sect5}
In this section, we evaluate the performance of our proposed solutions. We start with a description of the dataset and simulation environment. Afterwards, we explain the benchmark approaches and implementations that we use. Finally, we discuss the overall results.      




\begin{table}[ht]
\Huge
\centering
\caption{LIST OF PARAMETERS} \label{tablepara}
\begin{adjustbox}{width=0.47\textwidth}
\begin{tabular}{|l|c|}
\hline
Parameter & Value\\
\hline \hline
$P_{b}^{static}$ \cite{AftabHossain2018} & 18 Watts\\\hline
$P_{b}^{fixed}$ \cite{Debaillie2015}  & 139 Watts\\\hline
$P_{b}^{dynamic}$ \cite{Debaillie2015} & 742 Watts \\\hline
$\psi_{i}$ [Facebook, YouTube, Google, EcoSlice] & [1.2, 1.6, 1.4, 1]    \\\hline
$\delta_{i}$ [Facebook, YouTube, Google, EcoSlice] & [10, 1, 15, 11] ms \\\hline
Number of users per slice & [11-30] \\\hline
Users delay requirement $d_{u}^{\tau}$ & \begin{tabular}[c]{@{}l@{}}Facebook: [11-20]ms \\ YouTube: \ [6-17]ms\\ Google:  \ \ \ [16-25]ms \end{tabular} \\\hline
Loss function & MSE \\\hline
Learning rate $\alpha$ & 0.001 \\ \hline
$\beta$  & [5,1,0.8] \\\hline
Maximum episodes $J$& 1000\\\hline
$\varphi$ & 0.5  \\\hline
$M$ & 0.01 \\\hline

\end{tabular}
\end{adjustbox}
\end{table}

\subsection{Dataset and Simulation Setup}

\begin{figure*} 
    \centering
    \begin{subfigure}[h]{1\textwidth}
{\includegraphics[scale=0.35]{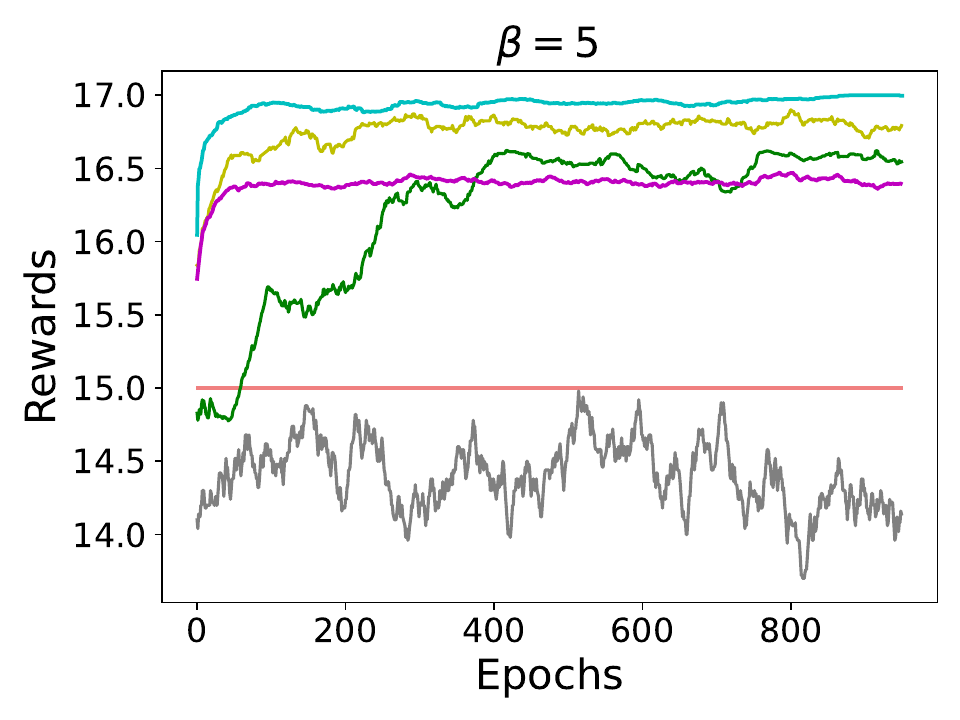}}
{\includegraphics[scale=0.35]{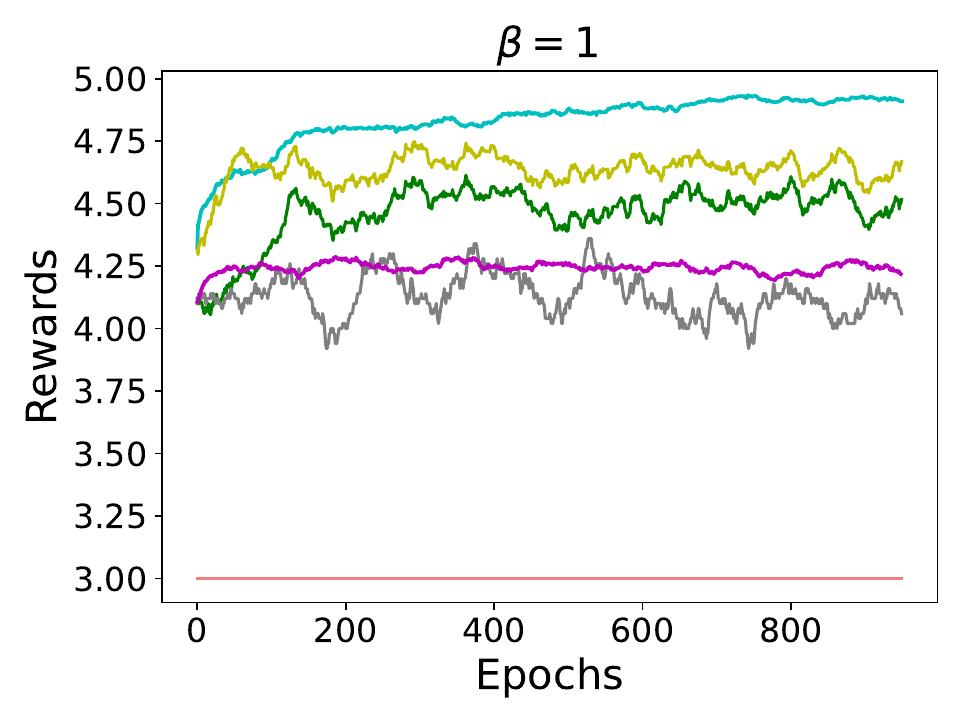}}
    {\includegraphics[scale=0.35]{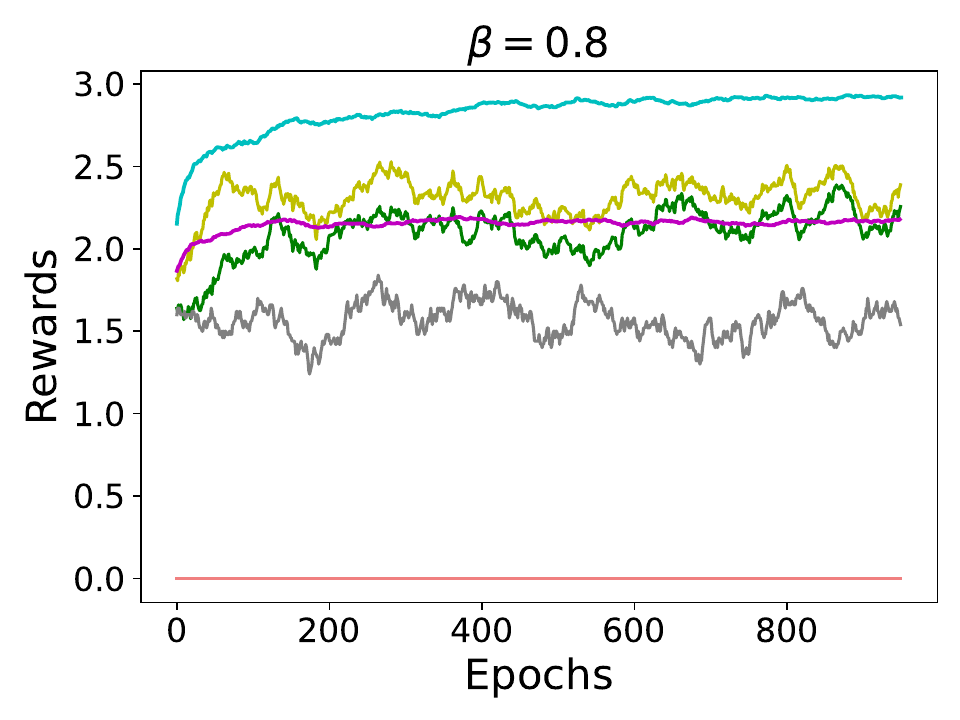}}
    \end{subfigure}
     \begin{subfigure}[h]{1\textwidth}
    \centering
    {\includegraphics[scale=0.65]{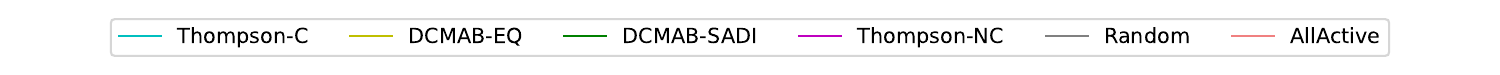}}
    \subcaption{Reward.}
    \end{subfigure}
    \begin{subfigure}[h]{1\textwidth}
    {\includegraphics[scale=0.35]{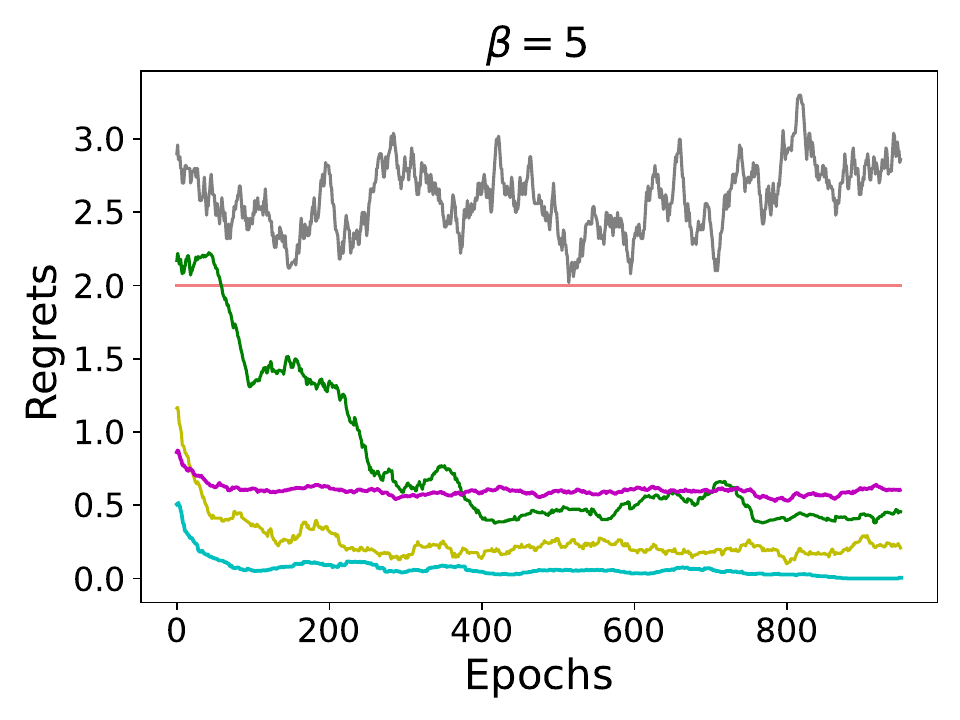}}
    {\includegraphics[scale=0.35]{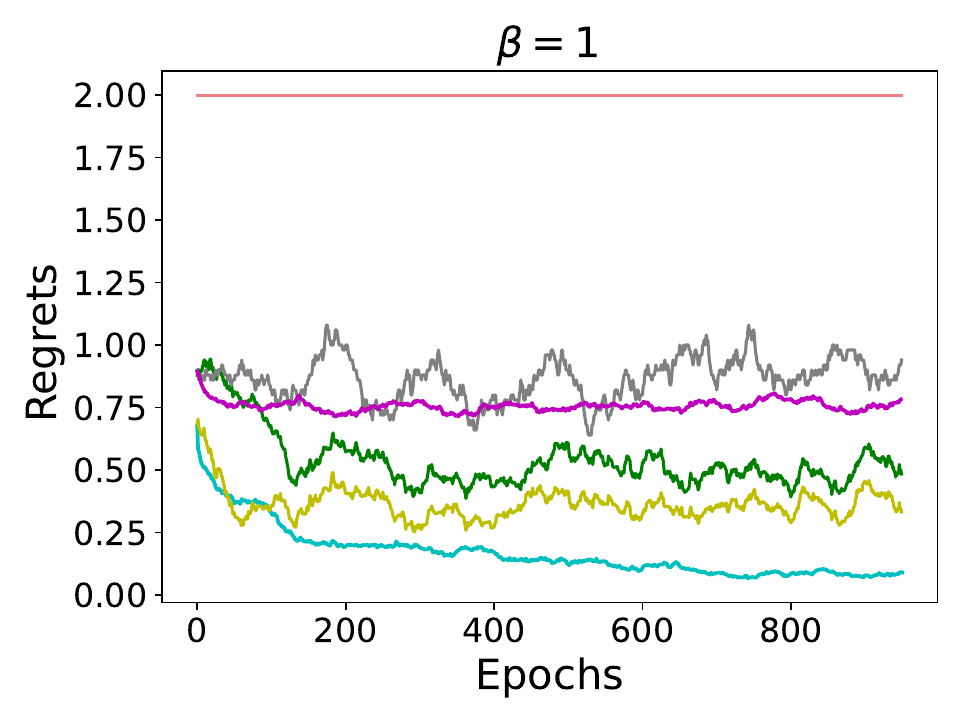}}
    {\includegraphics[scale=0.35]{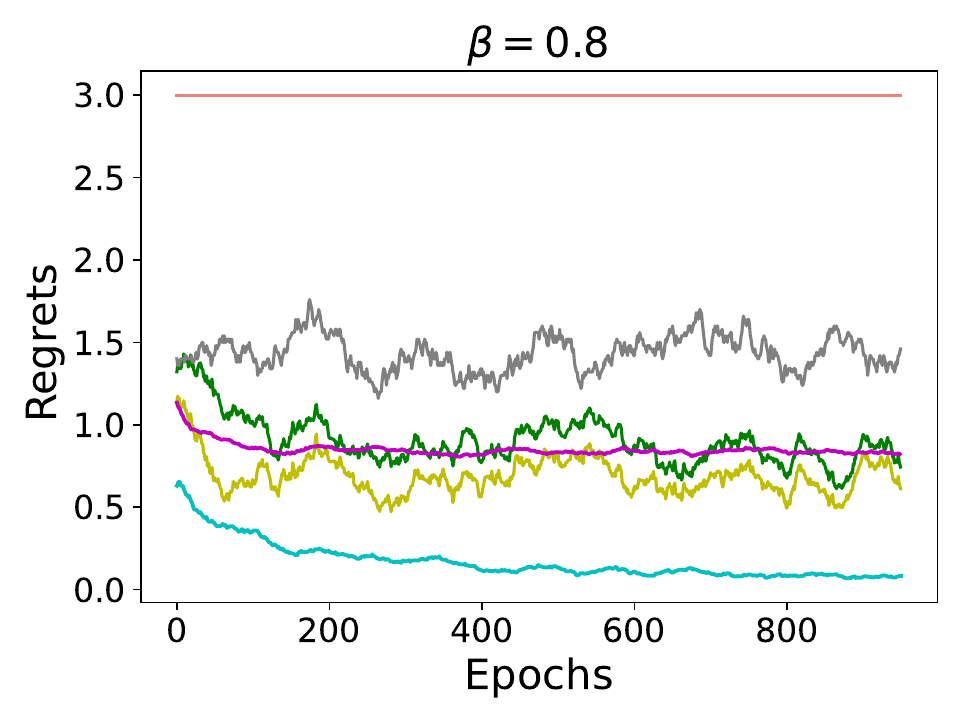}}
    \vspace{-0.1cm}
    \end{subfigure}
    \begin{subfigure}[h]{1\textwidth}
    \centering
    {\includegraphics[scale=0.65]{figures/legend.pdf}}
    \subcaption{Regret.}
    \end{subfigure}
    \caption{Reward and regret obtained for different $\beta$ values.}
    \label{rewardregret}
    \vspace{-0.15cm}
\end{figure*}

We evaluate the proposed solutions using a real-world dataset collected from the Orange 4G network, in Poitiers, France. The dataset includes mobile data traffic demand of different mobile applications at a base station level. We assume slice instances are deployed on an application-basis, i.e., one application maps to one slice instance. More precisely, Facebook, YouTube, and Google have been considered as three different types of slice instances attached to each base station. It is worth mentioning that those three different applications exhibit very different traffic demands, which we considerate appropriate for the simulation of network slicing~\cite{Hnin2022}. On the user side, we assume stochastic delay requirements of users for each application, as indicated in Table~\ref{tablepara}.

The granularity of the dataset is 10 minutes, for 10 days in May 2019. With this, for our simulation purposes, we consider $T$ is 10 days and SADI $\tau = 10 \ minutes$. Thus $T$ includes 1440 $\tau$. We analyse 10 base stations from the Poitiers city center, where different slice instances are associated. We then apply our proposed solutions using an action set where each action implies the activation/deactivation of one of the slices: $\left[\text{Facebook, YouTube, Google, EcoSlice}\right]$.

\begin{figure}[ht!] 
    \centering
    \includegraphics[scale=0.25, trim = {1cm 0cm 0cm 0cm}]{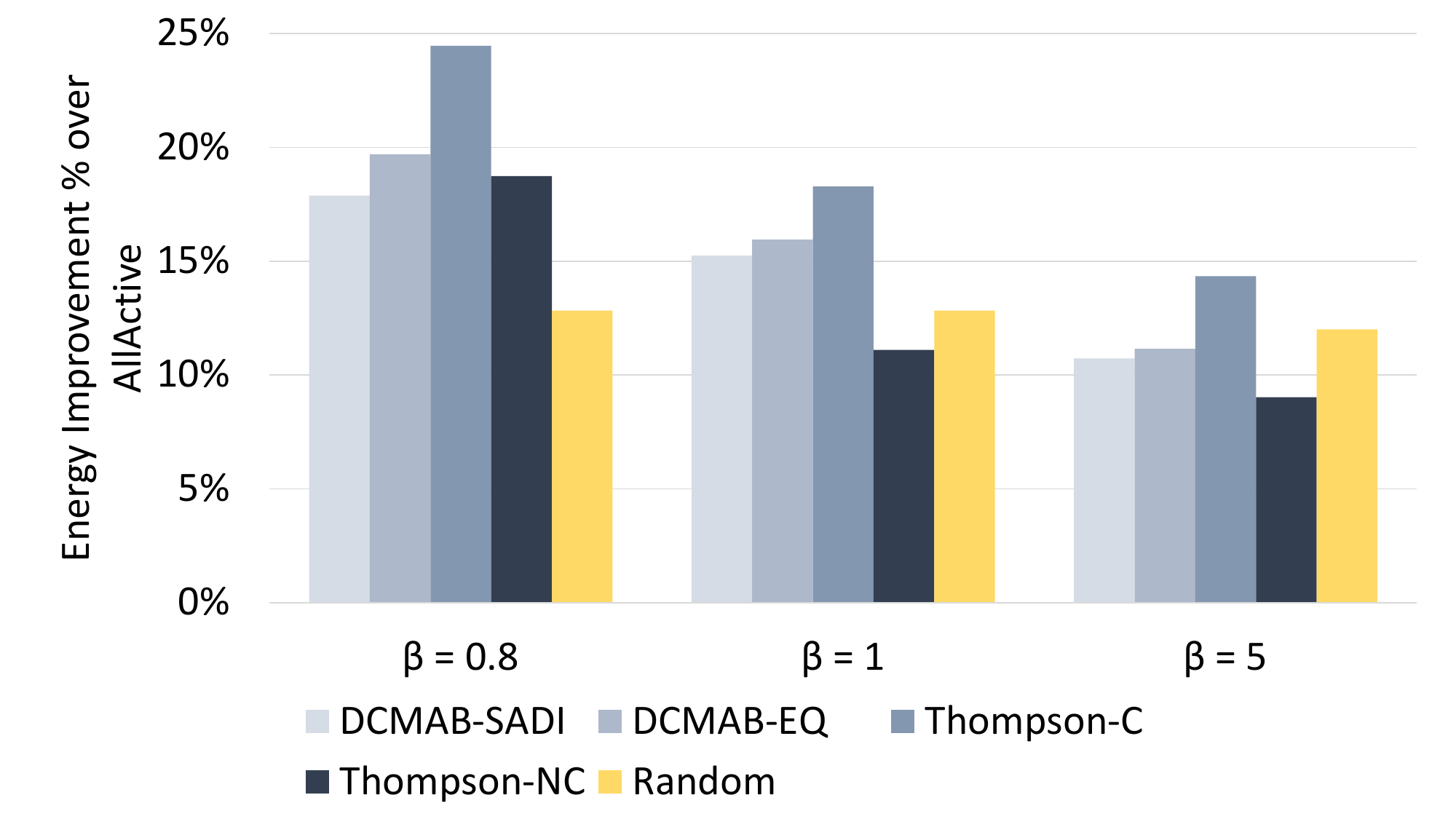}
    \caption{Energy improvement over AllActive}
    \vspace{-0.23cm}
    \label{energy}
\end{figure}

\begin{figure}[ht!] 
\vspace{-0.23cm}
    \centering
    \includegraphics[scale=0.25, trim = {1cm 0cm 0cm 0cm}]{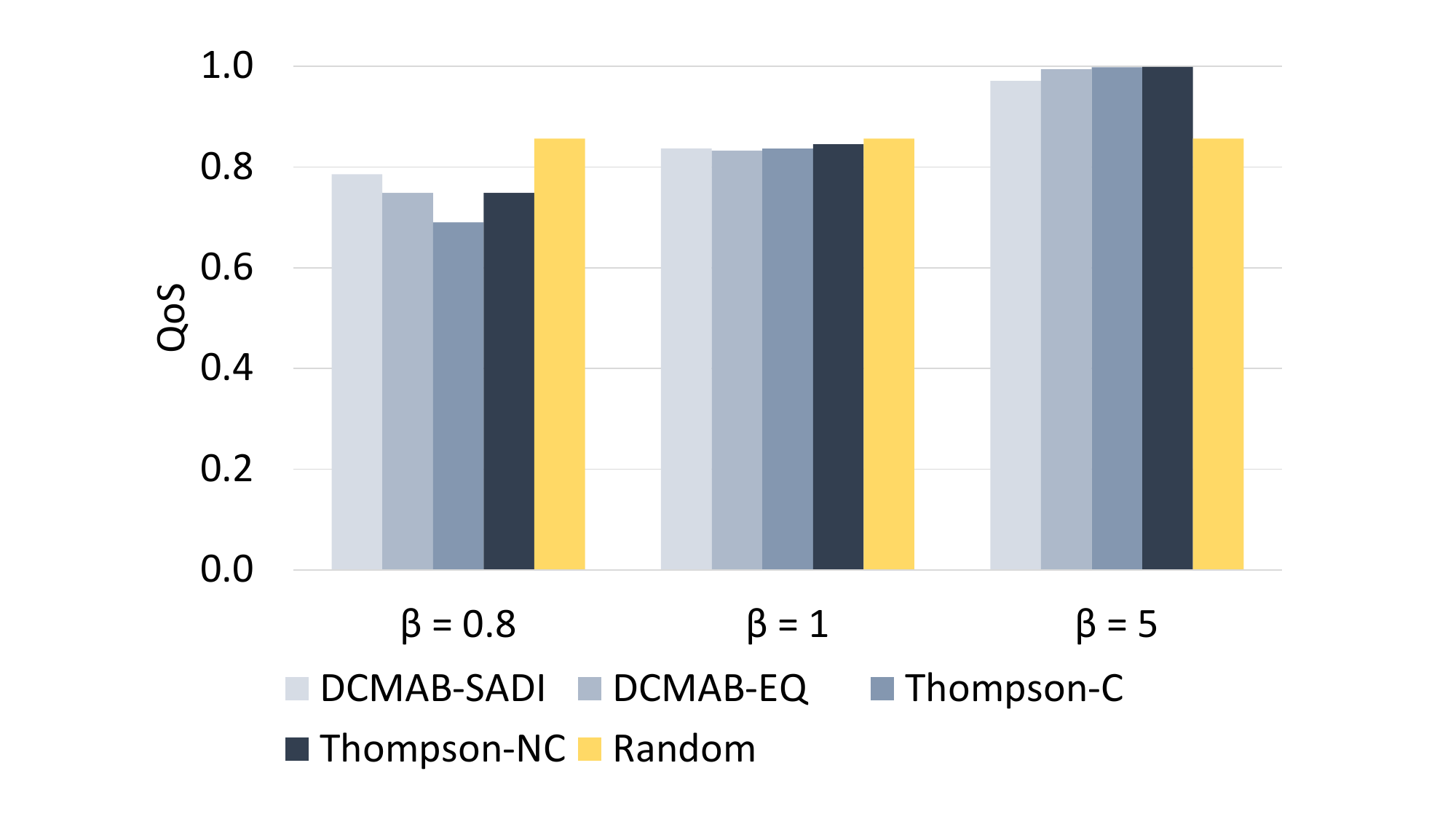}
    \caption{QoS based on different $\beta$ values}
    \label{qos}
\end{figure}

\subsection{Benchmarks and Implementation Setup}

We compare the performance of our proposed DCMAB and Thompson-C solutions with three counterparts:
Thompson Sampling Non-Contextual (Thompson-NC), AllActive and Random. Unlike Thompson-C, no state/context information is considered in Thompson-NC~\cite{Agrawal2017}. For AllActive, as the name implies it, all the slice instances are active. And a random action is selected at each iteration for the Random approach. 

For the implementation, we rely on the Pytorch framework for the DCMAB agents. All the models (i.e. DCMAB, Thompson-C, Thompson-NC, AllActive and Random) are implemented using a Python environment and trained on the high-performance Linux server provided by Digital Research Alliance of Canada. The DNN of a DCMAB agent is composed of three fully-connected layers of 100 neurons, followed by a RELU activation function for each. The detailed parameters are summarized in Table~\ref{tablepara}.

\subsection{Results}

\begin{figure*} 
    \centering
    \begin{subfigure}[h]{1\textwidth}
    \subfloat{\includegraphics[scale=0.28]{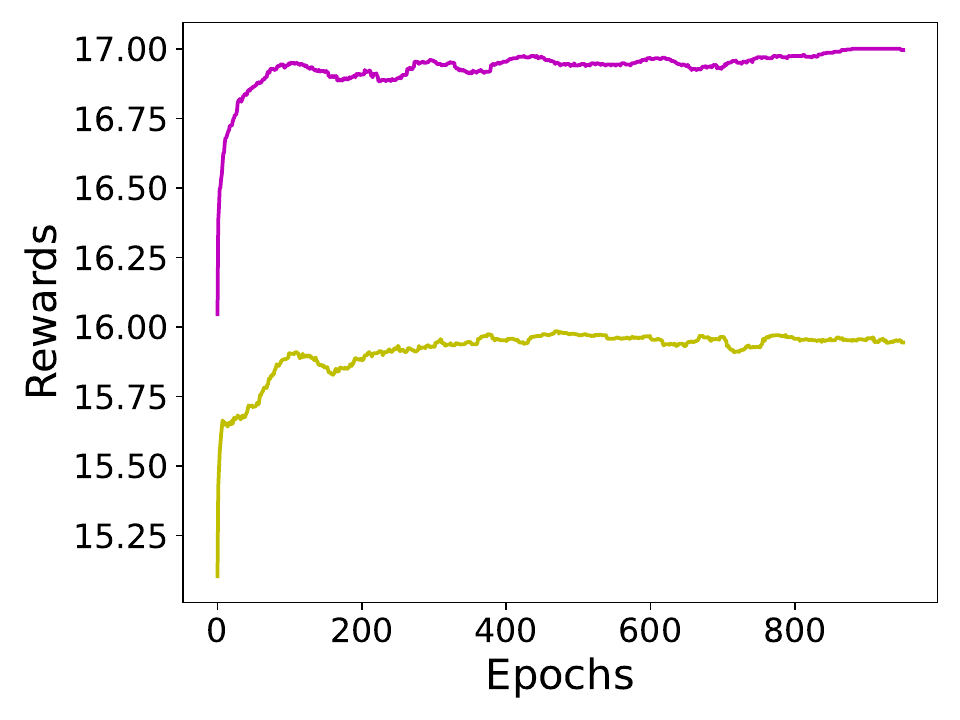}}
    \subfloat{\includegraphics[scale=0.28]{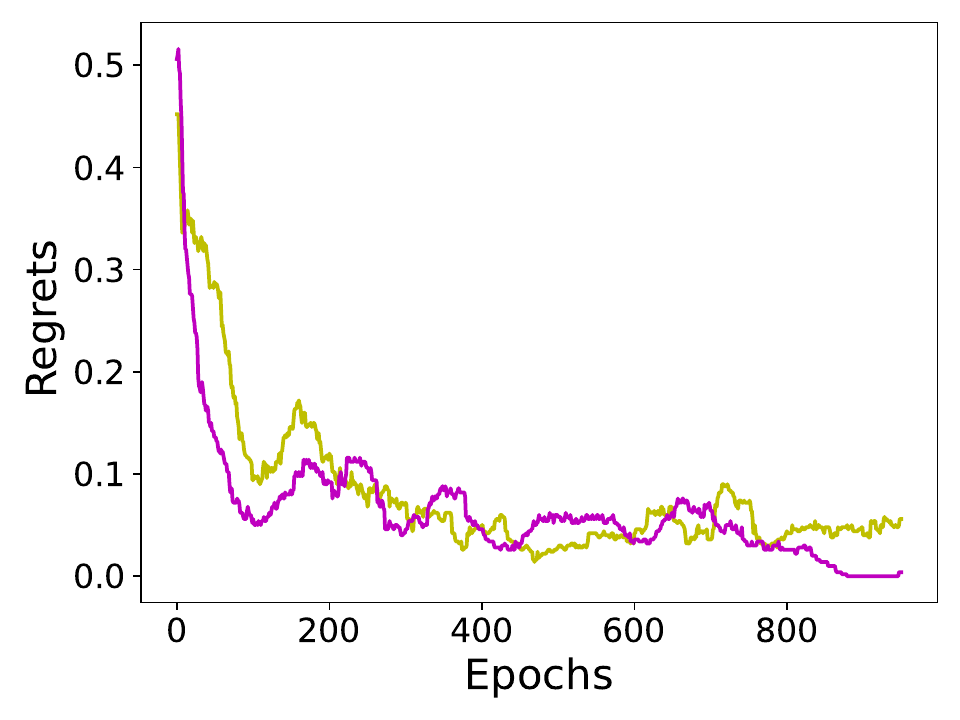}}
    \subfloat{\includegraphics[scale=0.28]{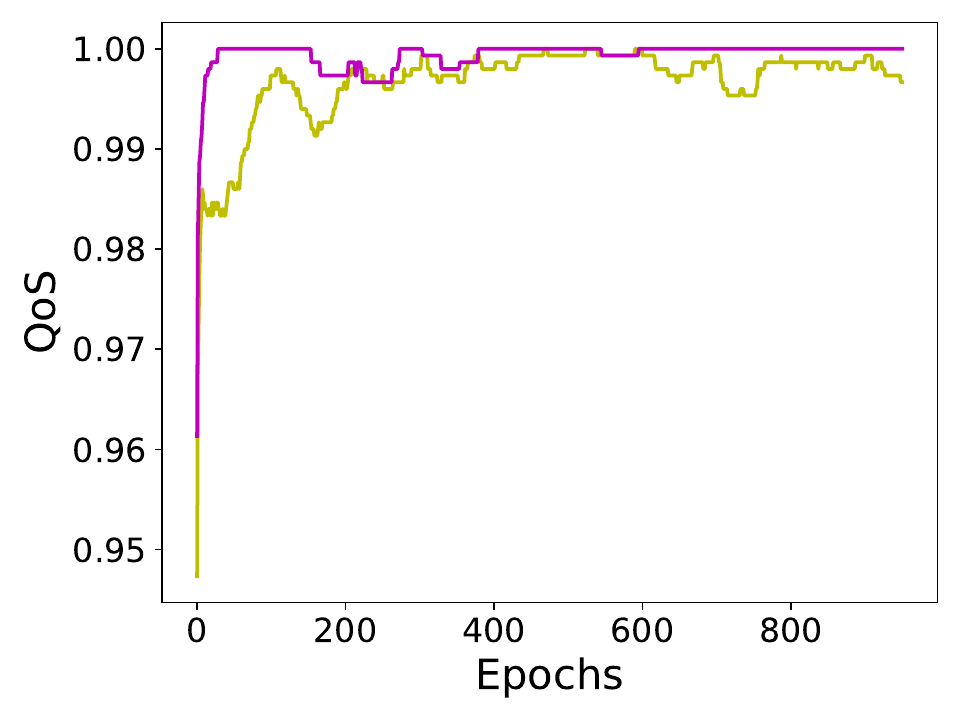}}
    \subfloat{\includegraphics[scale=0.28]{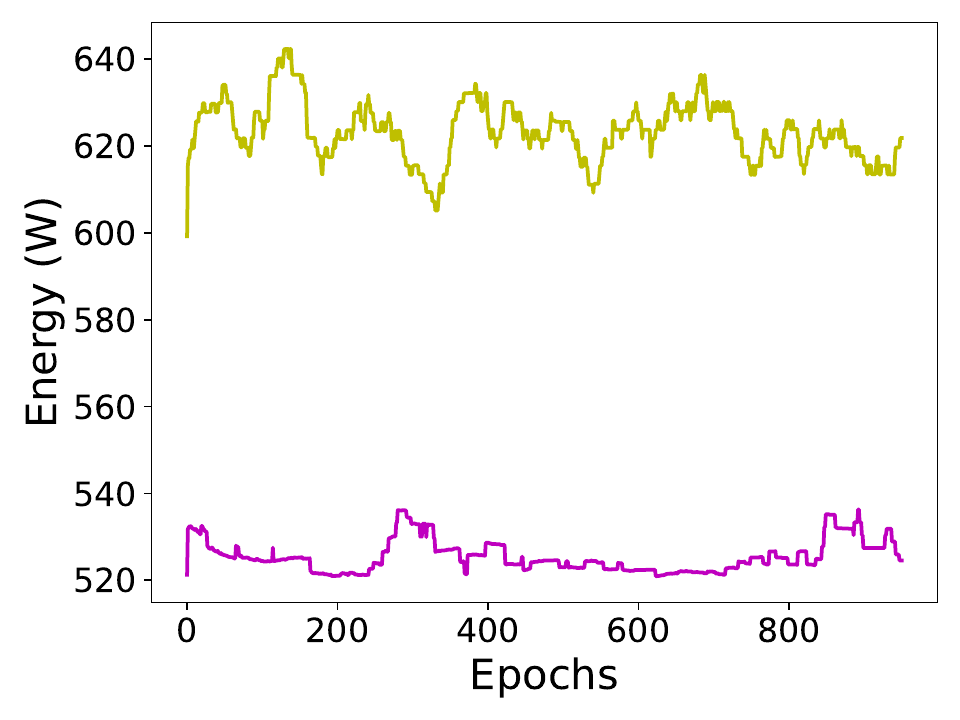}}
    \end{subfigure}
    \begin{subfigure}[h]{1\textwidth}
    \centering
    \subfloat{\includegraphics[scale=0.6]{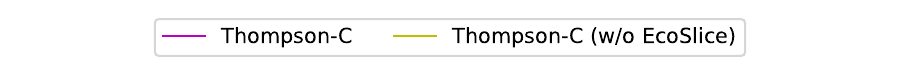}}
    \end{subfigure}
    \caption{Comparisons of Thompson-C Vs Thompson-C (w/o EcoSlice)}
    \label{woeco}
    
\end{figure*}

\subsubsection{Overall Agents Performances}

First of all, to fully comprehend the performance of our agents, we study the trends of reward, regret, QoS and energy for $\beta = [5, 1, 0.8]$. We note that agents focus on QoS when $\beta=5$, search for a trade-off when $\beta=1$, and stress on energy when $\beta=0.8$. Therewith, we compare the reward trends of our proposed solutions and their peers in Figure~\ref{rewardregret}. The curves are smoothed by averaging within a rolling window of 50 iterations. 

In Figure~\ref{rewardregret}a, the DCMAB-EQ and Thompson-C agents exhibit the superior reward, even significantly better than AllActive, followed by DCMAB-SADI and Thompson-NC. Random approaches failed our proposed solutions in every scenario. Noticeably, AllActive shows inferior performance to that of the other agents in general. We also perceive the same behavior for the regret trends for all the agents in Figure~\ref{rewardregret}b.

\subsubsection{Roles of Agents in Energy and QoS}
We then explicitly verify if our proposed solutions are feasible for energy optimization in RAN slicing by comparing them with the baselines. In this regard, as depicted in Figure~\ref{energy}, compared with the AllActive strategy (currently the standard approach), the energy improvement of Thompson-C is approximately 24\%, 18\% and 14\% respectively, for $\beta$ equals 0.8, 1 and 5. As expected, the energy gain deteriorates when $\beta$ value increases. The prior phenomenon holds for all the agents (except Random). Also, we saw a great deal of energy gains over AllActive for DCMAB-EQ, DCMAB-SADI, Thompson-NC and even Random as well. Overall, we notice the Thompson-C agent is quite superior to others in terms of energy gain. Conversely, Thompson-NC has slightly lower energy gains than the DCMAB agents for all $\beta$ values. 

We are all aware that optimizing energy consumption means compromising QoS to an extent. In this light, we visualize the QoS of all the agents in Figure~\ref{qos}. We note that Thompson-C exhibits a comparable performance with its fellows in terms of QoS, but a slightly lower performance when $\beta=0.8$. This is linked to Thompson-C showing highest energy gain at $\beta=0.8$. It is observed in Figure~\ref{qos} that our proposed solutions satisfy almost 100\% QoS at $\beta=5$ (slightly lower for DCMAB-SADI). Therewith, it is at $\beta=5$ where our agents make themselves stand out, as they deliver the same QoS as AllActive, while providing significant improvement in energy efficiency. Regardless of showing acceptable performance, the Thompson-NC agent shows lower performance than our proposed solutions, comforting our modelling choices. We can not help but stress that: \emph{i)} state-aware agents outperform the one in which context/state is not considered, and \emph{ii)} a DNN approach is constantly surpassed by a statistical approach. 

\subsubsection{Impact of EcoSlice}

To grasp the benefits EcoSlice, we examine in Figure~\ref{woeco} the performance of the Thompson-C agent with and witout an EcoSlice, in terms of reward, regret, QoS and energy utilization. We select the Thompson-C algorithm here as it shows the best results in most of the scenarios compared to its fellows. As we can observe, Thompson-C demonstrates better performance under the different metrics when compared to Thompson-C (w/o EcoSlice). Therefore, an EcoSlice significantly enhances the overall energy efficiency of the network by allowing operators to switch off the underutilized slice instances and yet ensure QoS. Without the assistance of an EcoSlice, one can not reach the level of energy efficiency that we have accomplished. 

\begin{figure}[ht!] 
    \centering
    \includegraphics[scale=0.25, trim = {0cm 0cm 0cm 0cm}]{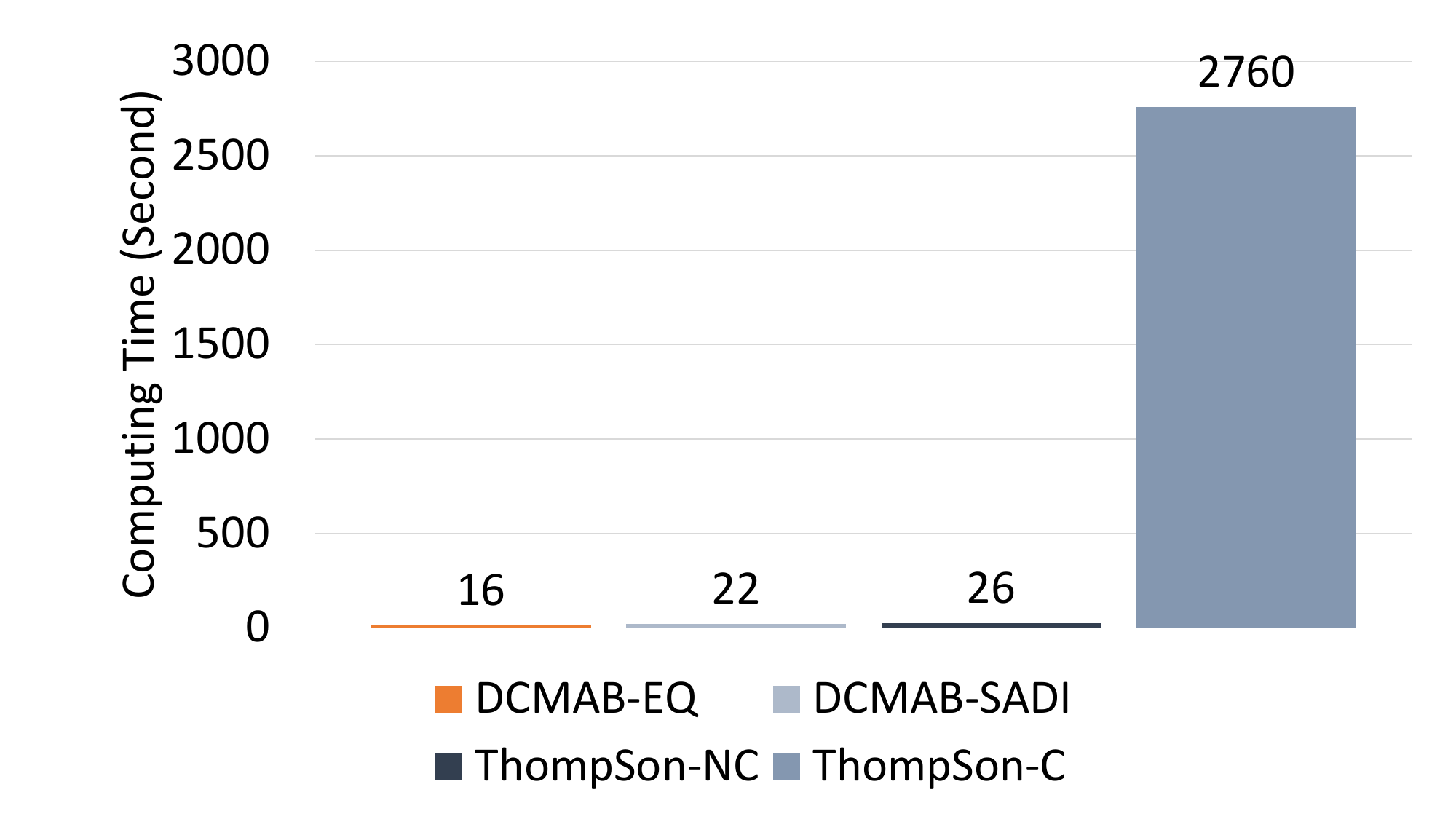}
    \caption{Computing time comparisons}
    \label{comput}
\end{figure} 


\subsubsection{Computing Time Comparisons}
The detailed computing time comparisons conducted on the  Digital Research Alliance of Canada servers are shown in Figure~\ref{comput}. Paying the price for its performance, Thompson-C takes much longer computing time than all the other agents. Despite Thompson-C being considered to be a better agent than DCMAB-EQ, it is not a good option for the real-time decision-making process, at least not with a SADI of 10 minutes, as we consider. On the other side, the DCMAB agents, who also outperformed the baselines in terms of energy efficiency, are on par with Thompson-NC in terms of computing power. There is no single answer here, and Thompson-C can be a favourable solution for a system without computing time constraints. However, all the solutions we proposed in this work offer avenues for operators to optimize the energy efficiency by slice activation/deactivation while controlling the impact on QoS. Operators can also opt between different design choices by varying $\beta$, based on their targets and limitations.



\begin{figure}[ht!] 
    \centering    \includegraphics[scale=0.4,  trim = {0cm 0cm 0cm 0cm}]{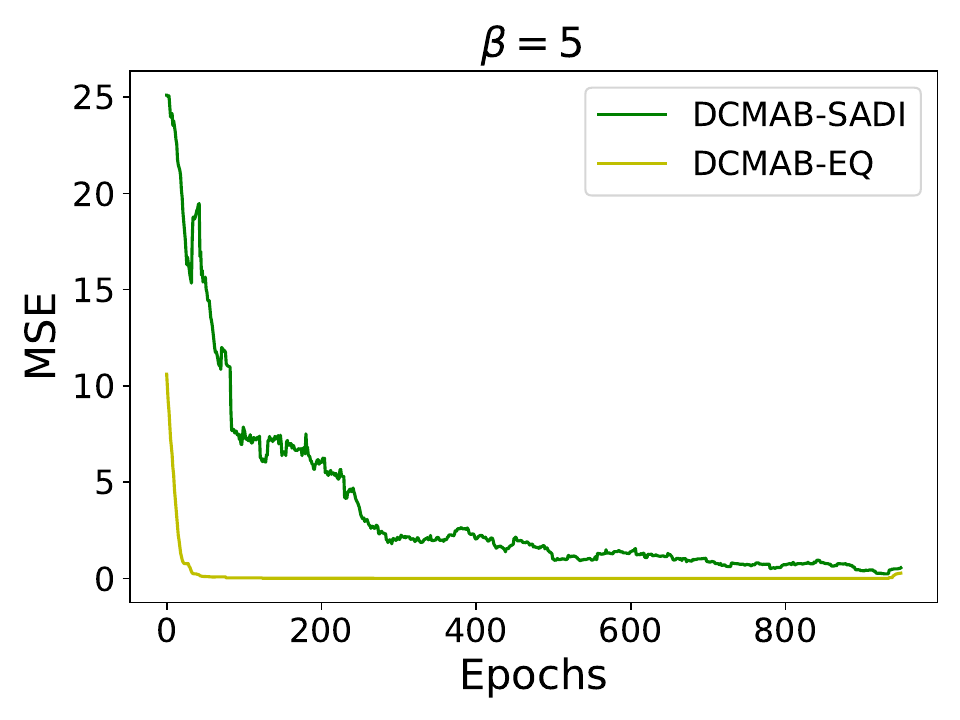}
    \caption{MSE comparisons of DCMAB-EQ Vs DCMAB-SADI}
    \label{mse}
    
\end{figure} 

\subsubsection{MSE of DCMAB} 
Last but not least, we also compare the MSE of the two DNN-based solutions, DCMAB-EQ and DCMAB-SADI, in Figure~\ref{mse}. As we can see, DCMAB-EQ has a stable training process with lower MSE than DCMAB-SADI. This explains the superior performance of DCMAB-EQ in the prior results, as DCMAB-EQ predicts better reward distribution of associated actions. In Figure~\ref{mse}, MSE results are shown for $\beta=5$ only, but we noticed similar behaviour for other tested values.

\section{Conclusion} \label{sect6}

In this paper, we focus on the slice activation/deactivation problem, to further enhance the energy efficiency in RAN slicing. To this end, we advocate the state-aware MAB approaches (i.e., DCMAB and Thompson-C), where an agent attempts to activate the optimal slice instances while providing guaranteed QoS. More than anything else, we investigate the important aspect of the compromise between energy consumption and user QoS. The results are derived based on a real-world datasets and demonstrate that the MAB approach in general, and DCMAB and Thompson-C in particular, are appropriate for the slice activation/deactivation problem. They significanlty alleviate the energy consumption at the base station level while ensuring a satisfactory QoS level. 


\bibliographystyle{IEEEtran}
\bibliography{IEEEabrv,References}

\end{document}